\documentclass[prl, twocolumn,superscriptaddress,preprintnumbers,a4paper,showpacs,floatfix]{revtex4}
\usepackage{graphicx}
\usepackage{color}
\usepackage{amssymb}
\usepackage{amsmath}
\usepackage{ulem}
\usepackage{multirow}
\usepackage{dcolumn}

\color[rgb]{0.000,0.000,0.000}

\begin{document}

\title{A valence bond liquid on the honeycomb lattice}
\author{Simon A. J. Kimber}
\email[Email of corresponding author:]{kimber@esrf.fr}
\affiliation{European Synchrotron Radiation Facility (ESRF), 6 rue Jules Horowitz, BP
220, 38043 Grenoble Cedex 9, France}
\author{I. I. Mazin}
\affiliation{Code 6393, Naval Research Laboratory, Washington, DC 20375, USA}
\author{Juan Shen}
\affiliation{Institut f\"ur Theoretische Physik, Goethe-Universit\"at Frankfurt, 60438
Frankfurt am Main, Germany}
\author{Harald O. Jeschke}
\affiliation{Institut f\"ur Theoretische Physik, Goethe-Universit\"at Frankfurt, 60438
Frankfurt am Main, Germany}
\author{Sergey V. Streltsov}
\affiliation{Institute of Metal Physics, S. Kovalevskaya Street 18, 620041 Ekaterinburg,
Russia}
\affiliation{Ural Federal University, Mira Street 19, 620002 Ekaterinburg, Russia}
\author{Dimitri N. Argyriou}
\affiliation{European Spallation Source ESS AB, Box 176, 22100, Lund, Sweden}
\affiliation{Department of Synchrotron Research, Lund University, Box 118, Lund, Sweden.}
\author{Roser Valent\'i}
\email[Email of corresponding author:]{valenti@itp.uni-frankfurt.de}
\affiliation{Institut f\"ur Theoretische Physik, Goethe-Universit\"at Frankfurt, 60438
Frankfurt am Main, Germany}
\author{Daniel I. Khomskii}
\affiliation{II. Physikalisches Institut, Universit\"at zu K\"oln, Z\"ulpicher Strasse
77, 50937 K\"oln, Germany}
\date{\today}
\date{\today}

\begin{abstract}
  The honeycomb lattice material Li$_2$RuO$_3$ \ undergoes a
  dimerization of Ru$^{4+}$ cations on cooling below $270^{\circ}$C,
  where the magnetic susceptibility vanishes. We use density functional theory calculations
  to show that this reflects the formation of a 'valence bond
  crystal', with a strong bond disproportionation. On warming, x-ray
  diffraction shows that discrete three-fold symmetry is regained on
  average, and the dimerization apparently disappears.
 In contrast, 
  local structural measurements using high-energy x-rays, show that
   disordered dimers survive at the nanoscale up to at least 650$^\circ$ C.
 The high temperature phase of Li$_2$RuO$_3$  is
  thus an example of a valence bond liquid, where thermal fluctuations
  drive resonance between different dimer coverages, a classic analogue 
  of the
  resonating valence bond state often discussed in connection with high T$_c$ cuprates.
\end{abstract}

\pacs{61.05.cp,61.05.fm,61.50.-f,71.15.Mb}

\maketitle

\noindent \textit{Introduction.---}Honeycomb layered systems are of
particular interest in solid state physics, due to their fascinating 
electronic, magnetic and superconducting properties. Recently,
honeycomb iridates A$_2$IrO$_3$ (A=Li,Na) have attracted a lot of attention
due to the presence of large spin-orbit coupling that was suggested to lead to interesting
topological properties~\cite{Shitade2009} and possibly to
unconventional Kitaev magnetic interactions~\cite{Jackeli1}. The latter
proposition, however, is presently under discussion~\cite{Coldea}. Analogues
containing $4d$ metals, where the spin-orbit coupling is much weaker,
might thus offer further insights into the iridates~\cite{Li2RhO3}.

A well known member of the $4d$ family is Li$_2$RuO$_3$, which is
 similar to Li$_2$RhO$_3$ and Na$_2$IrO$_3,$ but with
one more hole in the transition metal $t_{2g}$ bands.  In Li$_2$RuO$_3$ 
an intriguing
ground state is found below $\sim 270^{\circ}$C, where the Ru atoms
form structural dimers with a very strong disproportionation of the short and 
long bonds ($l_{\mathrm{l}}/l_{\mathrm{s}}\sim 1.2)$~\cite{Miura}. The
origin of this dimerization is controversial, and it has been unclear
to what extent this behavior is present in the corresponding $d^5$
compounds; existing experimental evidence suggests that
Na$_2$IrO$_3$~remains highly symmetric, with less than a 3\%~variation
in Ir-Ir distances \cite{Coldea}, and the same is probably true for
Li$_2$RhO$_3$ and Li$_2$IrO$_3$~\cite{Gretarsson2013,Li2RhO3}. On the
other hand, the relatively low quality of the samples does not allow
excluding a possible local structural dimerization without long range
order, in which case the average crystal structure remains symmetric,
but the electronic physics would be highly influenced by a local-scale
dimerization. To assess this possibility, first of all one needs a
clear microscopic understanding of the well established dimerization
in Li$_2$RuO$_3.$ The authors of the original work reporting
dimerization~\cite{Miura} later interpreted it as formation of
covalent bonds between a particular pair of Ru orbitals~\cite{Miura0},
and supported this conjecture by looking at the character of the wave
functions at two high-symmetry points in the Brillouin zone.

In this Letter, we show that the tendency to dimerization is local
and, as previously conjectured~\cite{Miura}, driven by covalency. Moreover, we find experimentally that
the dimerization locally
 survives well above the transition temperature, forming a
valence bond liquid (VBL). Approximately 1/3 of all Ru-Ru bonds are
dimerized at all temperatures and the dimer ordering is short ranged
with a correlation length of the order of a few nanometers at high
temperatures. The VBL we propose is not to be confused with the
so-called \textit{resonating} valence bond (RVB)
liquid~\cite{Anderson} that has received much attention in connection
with high $T_c$ cuprates; in the latter the many-body electron wave
function is a linear combination of the electronic states with all
possible spin singlets (called \textquotedblleft valence
bonds\textquotedblright ). In particular one can define a RVB state
with dimer singlets (short-range RVB)~\cite{Rokhsar1988} where the
\textit{resonating} nature is a result of the quantum fluctuations. In
the present case the valence bond liquid originates from existing
dimer patterns that \textit{resonate} due to thermal fluctuations and
could be described as the classical analog of short-range RVB.

Our density functional
theory (DFT) calculations strongly favor the formation of local
dimers with the experimentally observed long range order providing the lowest
energy. However, other mutual arrangements of Ru-Ru dimers have similar
energies, varying by $\approx$ 40~meV, while all dimerized patterns have
the energy much lower (by $\approx$ 150~meV) than the energy of the uniform state
without structural dimers. This explains the fact that the long range ordering of dimers
is destroyed at $T_c$ $\approx$ 270$^o$C ($\approx$ 47~meV), whereas dimers survive
up to much higher temperatures.
 This result  supports  the
concept of a valence bond liquid, and is consistent with our experimental observation that 
upon quenching the high-temperature phase to 50$^\circ$ C the dimerization is recovered. Finally,
by analyzing the total density of states (DOS), we conclude
that not only the orbital identified in Ref.~\onlinecite{Miura0}
contributes to the covalency of the dimerized bond \textit{via} direct
overlap, but also another orbital provides an additional contribution
\textit{via} an O-assisted hopping. These findings play a very
important role in understanding the microscopic physics of other $4d$
and $5d$ honeycomb oxides.

\begin{figure}[tb]
\begin{center}
\includegraphics[scale=0.25]{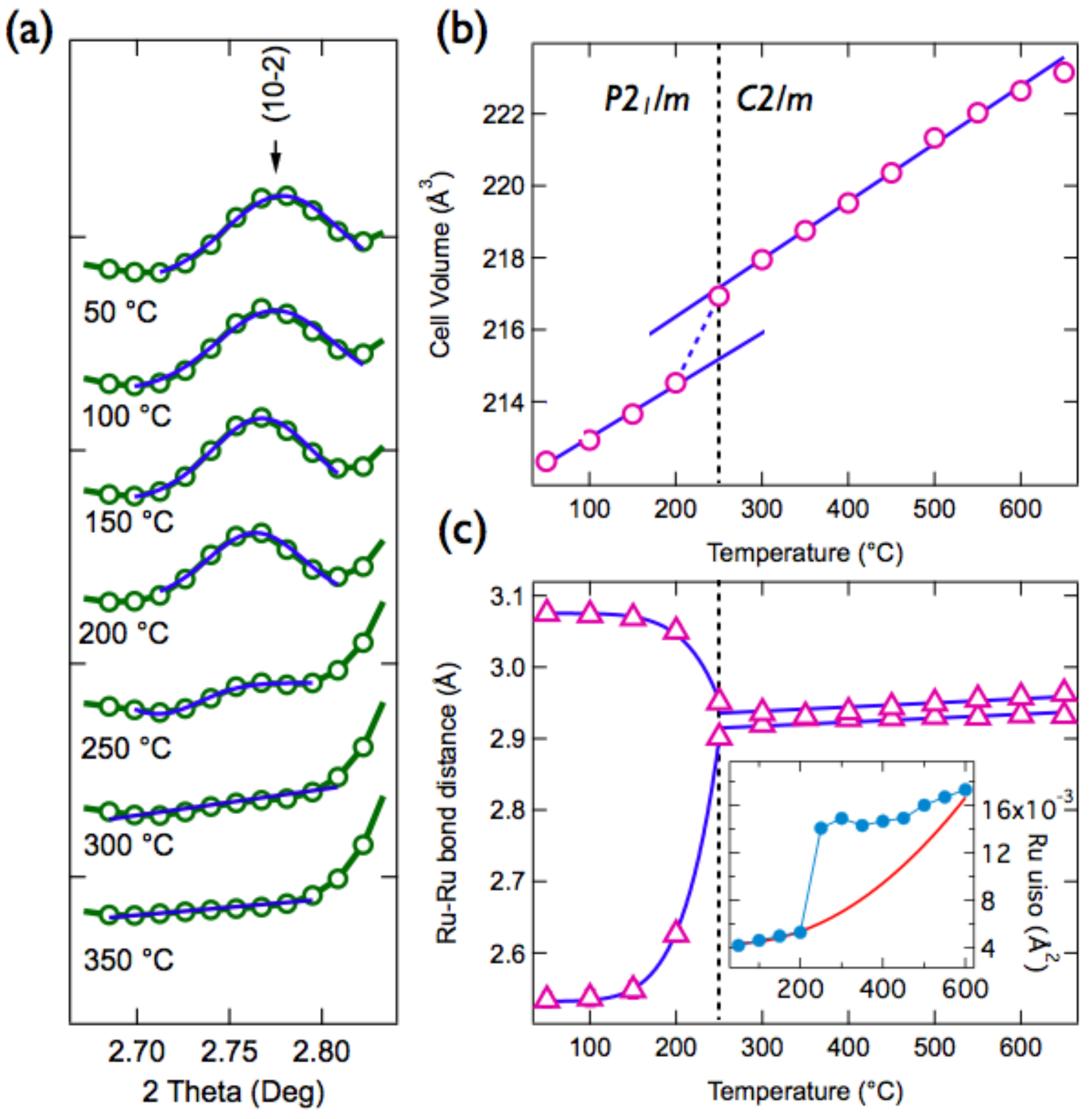}
\end{center}
\caption{(color online) (a) Selected region of the x-ray diffraction
  profile of Li$_2$RuO$_3$ as a function of temperature, showing the
  disappearance of the (10-2) reflection at the structural phase
  transition. The tail to higher angles originates from the intense
  (11-2) reflection. (b) Temperature dependence of the refined unit
  cell volume of Li$_2$RuO$_3$. (c) Temperature dependence of the
  Ru-Ru bond distance extracted from Rietveld refinement against the
  powder x-ray diffraction data, the inset shows the refined Ru atomic
  displacement parameter with an extrapolated Debye dependence (red line).}
\label{Powder}
\end{figure}

\begin{figure}[tbp]
\begin{center}
\includegraphics[scale=0.06]{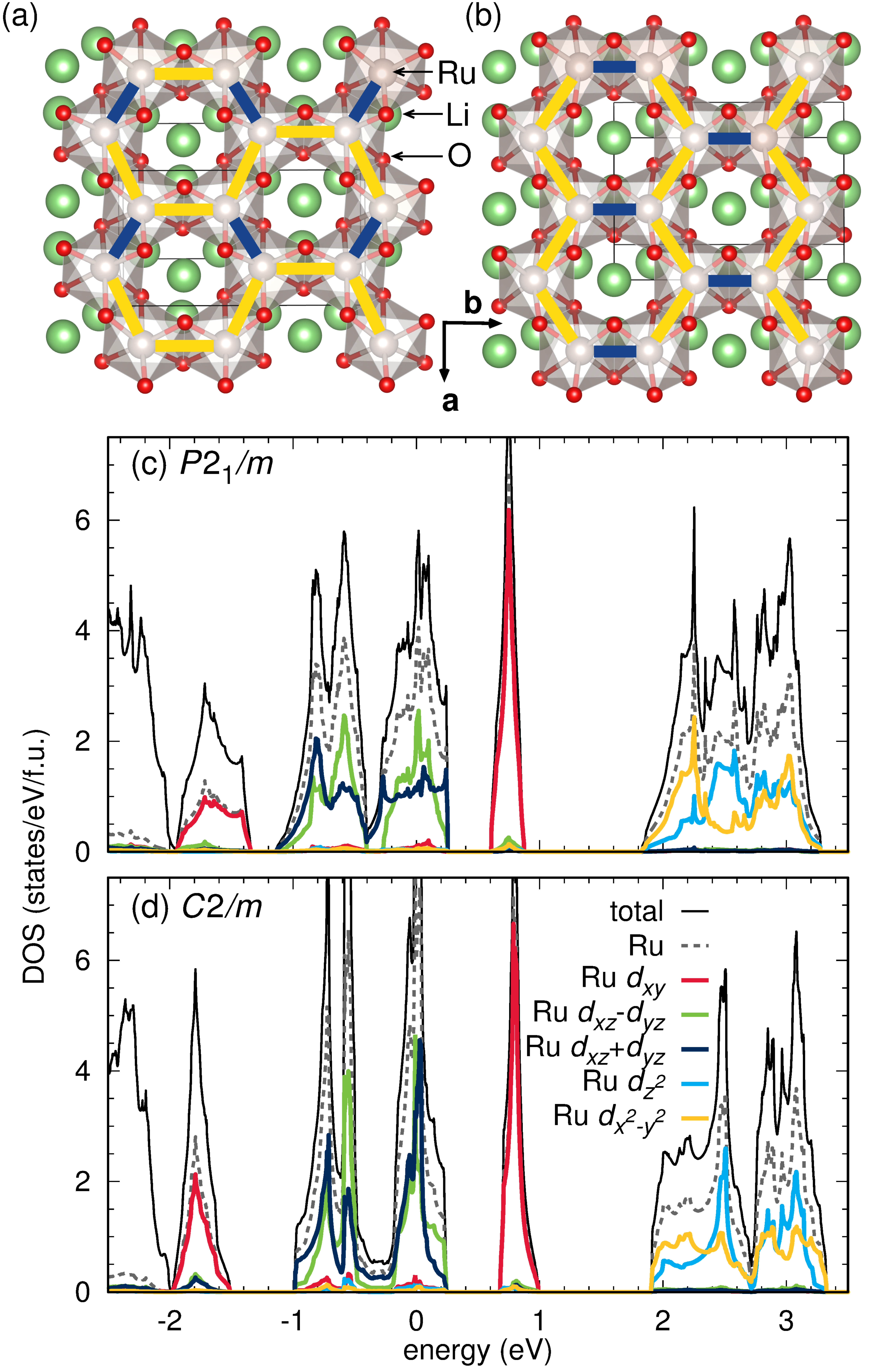}
\end{center}
\caption{(color online) Armchair (a) and parallel (b) dimerized
  structures of Li$_2$RuO$_3$ investigated in this work. Shorter
  (dimerized) Ru-Ru bonds are marked with a dark bar, longer bonds
  with a light bar. The Ru DOS for these two low energy
  structures are shown in panels (c) and (d) respectively. }
\label{STRUCT_DOS}
\end{figure}

\textit{Experiment.-} We synthesized a ceramic sample of
Li$_2$RuO$_3$, which was characterized by both neutron~\cite{Kimber}
and synchrotron x-ray powder diffraction. Both methods gave results
consistent with the dimerized $P\,2_1/m$ structure reported in Ref.~\onlinecite{Miura}
 at room temperature. Diffraction
experiments were performed using the ID15B beamline at the ESRF,
Grenoble. A wavelength of 0.1422~{\AA} was used and the scattered
x-rays were detected by a Mar345 image plate. Two detector distances
were used at each temperature, such that data suitable for both
Rietveld~\cite{Larson}~and pair distribution function analysis were
collected (see Supplementary Information (SI)). At room temperature, the dimerized structure is
evidenced by the presences of (\textit{h+k=odd}) reflections which
break \textit{C}-centering (Fig.~\ref{Powder}(a)). The corresponding
difference in Ru-Ru distances is pronounced, with 1/3 short bonds~\cite{comment_bonds}
(2.55~{\AA})~and 2/3 long ($\sim $3.1~{\AA}) (see Fig.~\ref{STRUCT_DOS}
(a)). Upon warming toward the transition, the (\textit{h+k=odd})
reflections lose intensity, merging into the background above
$250^{\circ}$C. The refined Ru-Ru distances converge to a single value
of $\sim $2.95~{\AA}, and at $350^{\circ}$C, excellent fits to the data
{were} obtained using the undistorted $C\,2/m$~structure (see SI).
The honeycomb layers are as symmetric as those seen~\cite{Kimber,
  Coldea}~in the metallic honeycomb ruthenate Ag$_3$LiRu$_2$O$_{6}$~or
Na$_2$IrO$_3$. {Although the bond lengths converge in a manner
  indicative of a displacive phase transition (Fig.~\ref{Powder}~(c)),
  a small volume anomaly is also observed (Fig.~\ref{Powder}~(b)),
  with a slight expansion upon entering the high temperature phase.}
Furthermore, a sharp increase in the Ru atomic displacement parameter
is observed, which implies an increase in disorder beyond that
expected in a simple Debye model (inset
Fig. ~\ref{Powder}~(c)).

Refinements against Bragg intensities are sensitive only to the
average crystallographic structure. We therefore performed pair
distribution function (PDF) analysis, which is sensitive to local
order as it includes the diffuse scattering signal. Using Igor Pro
based software (iPDF) developed by one of the authors (SAJK), we
corrected the raw diffraction data for background, incoherent
inelastic scattering, and the atomic form factors (see SI). To
visualize any short-range disorder, we Fourier transformed the
structure factors into real space using:
$G(r)=\frac{2}{\pi}\int_{0}^{\infty }Q[S(Q)-1]\sin (Qr)dQ$. Models
were fitted to the PDF data using the PDFgui package~\cite{farrow}.

\begin{figure}[tb]
\begin{center}
\includegraphics[scale=0.28]{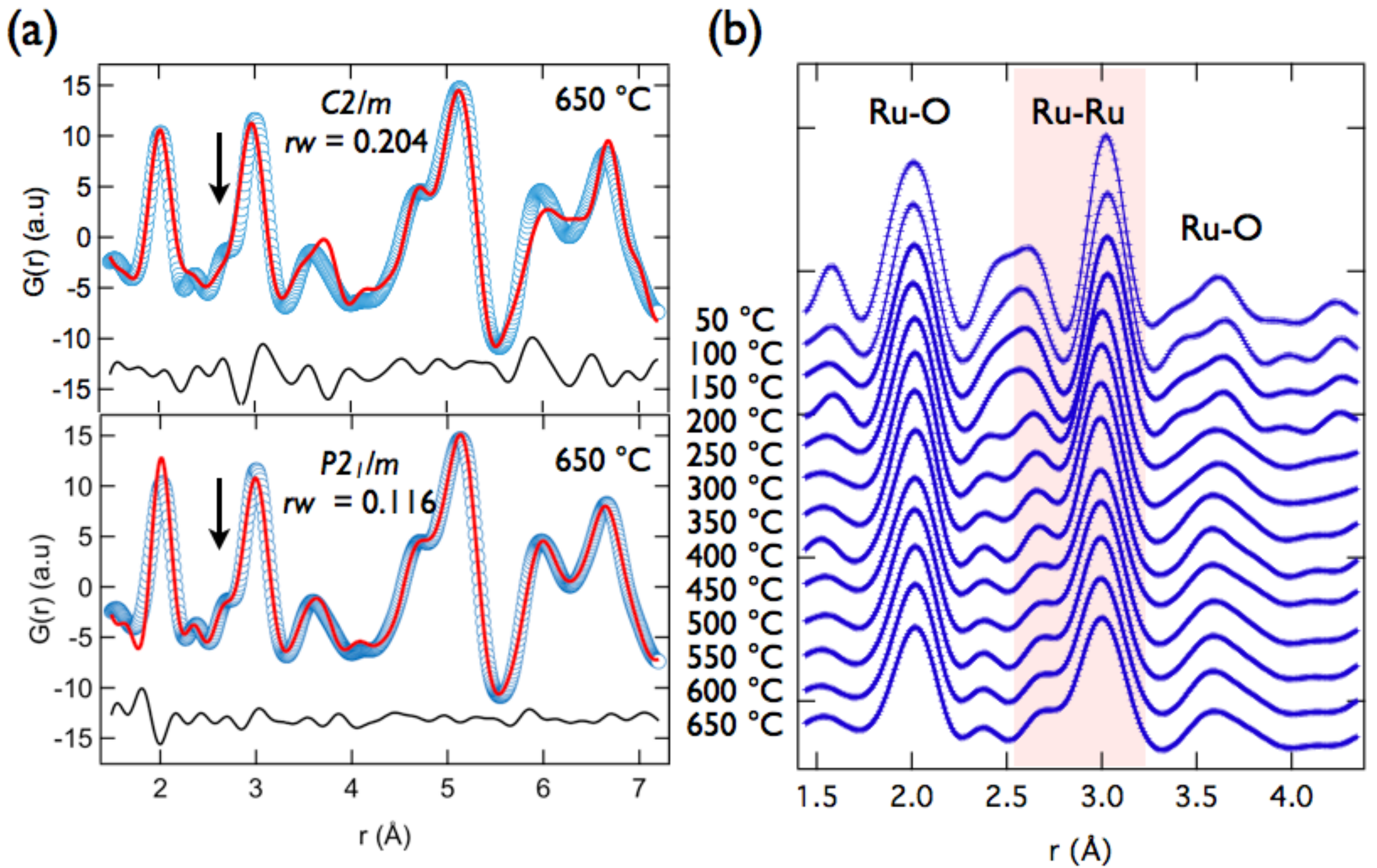}
\end{center}
\caption{(color online) a) Results of fitting the $C\,2/m$ (top) and
  $P\,2_1/m$ (bottom) structures to the pair distribution function
  (PDF) at $650^{\circ}$C (open blue circles: experiment; solid red
  lines: model calculations). The observed-calculated profile is the
  offset black line; b) Temperature evolution of the PDF.  The region
  containing the Ru-Ru bonds is highlighted and other important
  distances are also indicated. }
\label{PDF}
\end{figure}
\begin{figure}[tb]
\begin{center}
\includegraphics[scale=0.45]{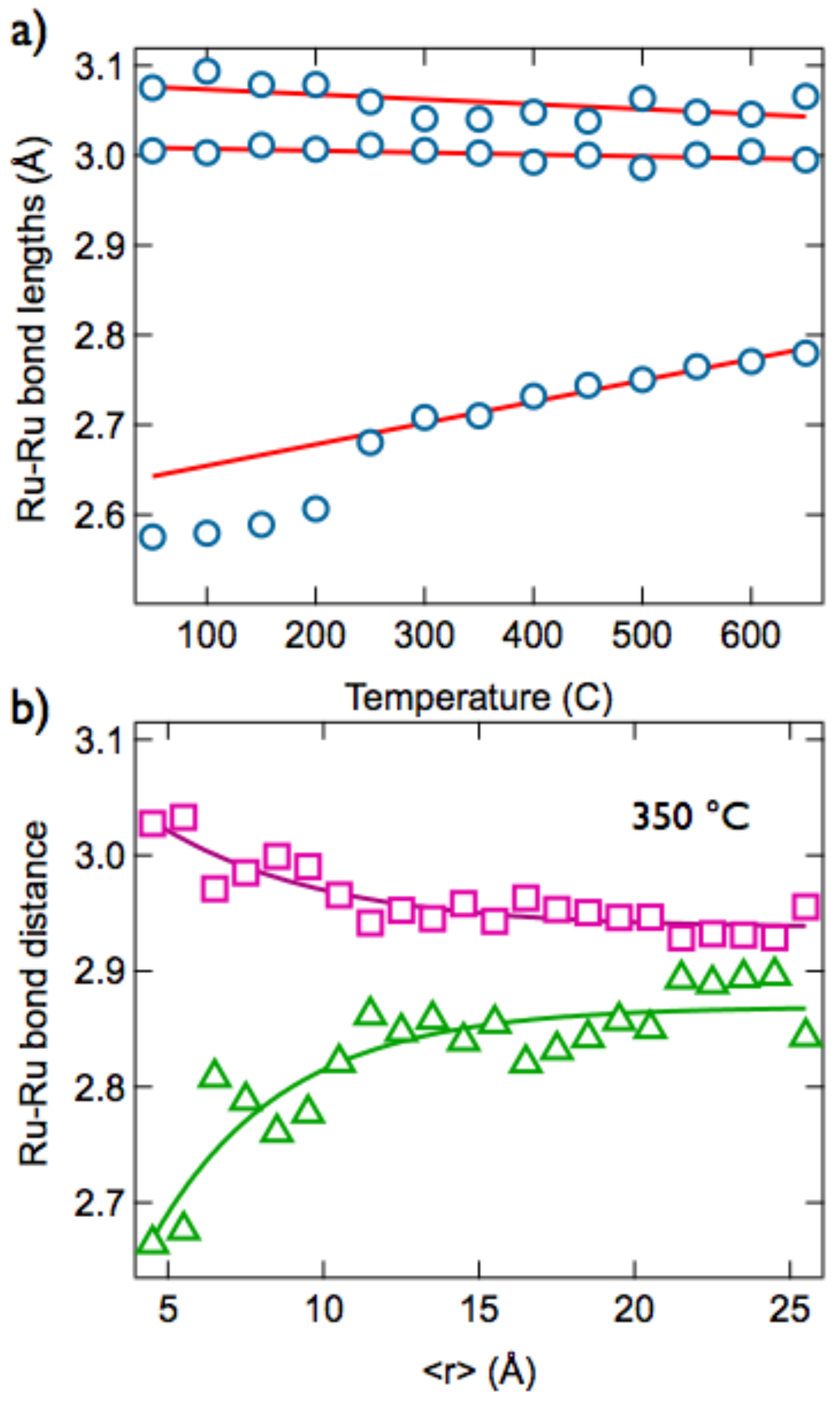}
\end{center}
\caption{(color online) (a) Temperature dependence of the fitted Ru-Ru
  distances from the PDF analysis of
  Li$_2$RuO$_3$ in the range 1.5 $<$ \textit{r} $<$ 7.25~{\AA}. Error
  bars obtained by directly fitting Gaussians to the data shown in
  Fig. 2 (b) are smaller than the markers; (b) \textit{r}-dependence of
  the Ru-Ru distances extracted from model fits to the PDF at
  $350^{\circ}$C as described in the text.  The two inequivalent
  longer distances have been averaged for plotting.}
\label{DIST}
\end{figure}

Deep within the dimerized phase of Li$_2$RuO$_3$, we obtained good
fits to the pair distribution functions using the $P\,2_1/m$~crystal
structure described above. However, upon examining the data at our
highest temperature ($650^{\circ}$C), we found that the fit of the
average $C\,2/m$ model was poor (\textit{rw}=0.204), 
when refined against the PDF in the range $1.5 < 5 < 7.25$~{\AA}
(Fig.~\ref{PDF}~(a)). In particular, the model does not reproduce a
prominent peak (arrowed) at 2.68~{\AA}. However, when we used the
dimerized model which describes the low-temperature phase, this peak,
and indeed the whole \textit{r}-range was well-fitted. The quality of
the fit was nearly twice as good ($rw=0.116$), and the 2.68~{\AA} peak
was due to Ru-Ru dimerization surviving at high
temperatures. Furthermore, the excellent agreement of the
$P\,2_1/m$~structure shows that $l_{\mathrm{l}}/l_{\mathrm{s}}$ is
approximately conserved. The PDFs are shown in Fig.~\ref{PDF}~(b). The
peak corresponding to the short Ru-Ru distance is highlighted for the
whole temperature region. Although there is a small shift of this peak
to larger \textit{r}-values upon heating, the converging bond
distances observed by Rietveld analysis (Fig.~\ref{Powder}~(c)) are
clearly not observed on the local length scale. When we repeated
refinements of the $P\,2_1/m$~structure on the local length scale for
the whole temperature range, we obtained the results shown in
Fig.~\ref{DIST}(a) for the Ru-Ru distances. The temperature variation
is almost linear with only a small anomaly at the $P\,2_1/m
\rightarrow C\,2/m$~transition. Extrapolating our data suggests that
the short-range dimerization would persist up to at least
$1400^{\circ}$C, $i.e$. well beyond chemical decomposition. This
energy scale ($\theta \sim $140 meV), which in a mean-field picture
corresponds to the dimerization energy, is rather large for magnetic
exchange interactions, although consistent with metal-metal covalent
bonding~\cite{Miura0,Kimber2}.  The energy scale of the ordering
transition, $\sim 47$ meV, on the same level of approximation, should
correspond to the inter-dimer interaction.

Having shown that the low temperature $P\,2_1/m$ structure provides a
reasonable model for the short-range dimer correlations at high
temperatures, we then proceeded to investigate the length scale of
this order. We performed so-called box-car refinement of this
structure against the PDF at $350^{\circ}$C. We used an 8.5~{\AA} box
and stepped \textit{r}-min from 1.5 to 32.5~{\AA} in 2.5~{\AA}
steps. The so refined Ru-Ru distances are shown in
Fig.~\ref{DIST}(b). We find that the difference between the short and
long bonds is progressively lost as the length scale increases
demonstrating that the dimer order at $T=350^\circ$C has a length
scale of $\sim 1.5$~nm (about two unit cells). At even higher
temperatures we found that the ordering of dimers vanishes beyond the
first coordination sphere. Thus, at $T\gg T_c=270^\circ$C there is
little correlation between individual dimers, but $\sim 1/3$ of all
Ru-Ru bonds remain dimerized. 

\textit{Theory.-} The origin of the room-temperature dimerization has
been discussed differently, depending on the starting point,
$i.e$. localized versus itinerant description. Jackeli and
Khomskii~\cite{Jackeli2} considered a localized picture and argued
that dimerization is controlled by spin physics, essentially,
promoting singlet formation on the short bonds. A drastic reduction of
the paramagnetic spin susceptibility below the structural transition
is consistent with this
concept. On the other hand, an  itinerant scenario was put
forward in Ref.~\onlinecite{Miura0}. These authors suggested that the driving
force of dimerization is covalent bonding. 
Both electrons of Ru$^{4+}$ can contribute to  dimer formation.   Indeed, we observe in
our electronic structure calculations that
for each bond there is one pair of Ru $4d$ orbitals that can overlap
directly (in an appropriately chosen local coordinate system, these
could be selected as $xy$ orbitals), with the amplitude $t_{dd\sigma
},$ and two that can overlap indirectly through two bridging oxygens
with the amplitude $\pm t_{pd\pi }^{2}/(E_{d}-E_{p}).$ Details and
visual illustrations can be found, for instance, in
Ref.~\onlinecite{Mazin}. One can
also construct linear combinations, $u_+=(xz+yz)/\sqrt{2}$
and $u_-=(xz-yz)/\sqrt{2}$. Both of them provide the same hoppings
via bridging oxygens, but only the former describes the direct hopping
($t_{dd\pi }$). For really short dimers (the low-T phase) the direct hopping dominates,
although it is largely offset by the indirect hopping, and 
as a result the bonding-antibonding
splitting between the $u+$ orbitals is larger than between
$u_-, $ but both are much smaller than the splitting
due to the $xy$ orbitals. For longer dimers (the high-T phase) $t_{dd\pi }$
becomes smaller, which results in smaller and less differentiated splitting
between the $u_\pm$ bands, ``unbounding'' the second hole and freeing it to respond
to external fields. 

To analyze this picture, we have performed full structural relaxations
of Li$_2$RuO$_3$ within DFT as implemented in the VASP
code~\cite{VASP}. The final energies were computed using an
all-electron method\cite{FPLO}. We considered a large sampling of
initial structures, including the experimentally reported structures
 and found several energetically
favorable structures; the simplest two  are shown in
Figs.~\ref{STRUCT_DOS} (a) and (b). The main results of our calculations
are that (1) the calculated ground state corresponds to a strong bond
disproportionation ($l_{\mathrm{l}}/l_{\mathrm{s}}\sim 1.2),$ in
quantitative agreement with the experiment; (2) The disproportionation
was only slightly stronger when magnetization of Ru is included; (3)
Examination of the DOS (Figs.~\ref{STRUCT_DOS} (c)
and (d)) confirms that the directly overlapping $xy$ orbitals form a
very strong covalent bond (bonding-antibonding splitting of more than
2 eV). One of the two holes of Ru residing in the $t_{2g}$ band is
occupying this {\it anti}bonding state, with a substantial energy gain;
(4) The $u_-$ orbital also contributes to the total
covalency, albeit considerably less.  The corresponding bonding-antibonding splitting is
about 0.7 eV and the second hole takes advantage of this fact. Moreover,
this contribution is only weakly dependent on the bond length, and
therefore its contribution to dimerization is much smaller than to
covalency in general; (5) Examination of the Ru effective moment in
spin-polarized calculation shows
that it is at most spin 1/2 and never 1; this indicates that at
least one electron spin is quenched. The second, $u_\pm$, spin may or
may not be quenched depending on temperature. Our calculations 
find rather small  energy differences between ferro- and antiferromagnetic
arrangement of the $u_\pm$ spins, so  when they
are unbound and their covalent bond is broken at high temperature,
they behave magnetically as nearly free spin 1/2 electrons. Indeed the
difference between the experimental susceptibility at low temperatures
(spin gap) and at high temperatures~\cite{Miura0} is consistent with this
scenario. Finally, (6) all investigated long range orders of the
structural dimers are energetically strongly favorable when compared
with the undimerized structure. The differences among the dimerized
structures are of the order of 40~meV ($\sim $ 450 K), comparable with
the ordering transition temperature. On the other hand, the
(optimized) uniform  structure is 155~meV ($\sim$ 1800K) above the
ground state structure, 
which explains why dimers themselves survive  well above $T_c$.

\noindent \textit{Discussion.-} Our experimental results combined with
theoretical calculations render the following picture: Ru-Ru bonds in
Li$_2 $RuO$_3$ have a very strong tendency to form local dimers with
covalent bonds \textit{via} direct overlap of Ru $4d$ orbitals. The
structural transition at $\sim 270^{\circ}$C is of the order-disorder
type: the dimers at $T \lesssim 270^{\circ}$C do not disappear at higher temperatures, nor does
their concentration (1/3 of all bonds) change. Dimer-dimer
interaction, presumably of elastic origin (as evidenced by the fact
that the disordered phase has a larger $a$ lattice parameter (see SI),
due to a lack of proper dimer packing), is much weaker and responsible
for mutual ordering of the dimers in the observed \textquotedblleft
armchair\textquotedblright\ structure.~\cite{Jackeli2} In the high temperature phase
there is no ordering of dimers at a length scale $\gtrsim 1.5$~nm, or
2-3 lattice parameters. The ordering temperature is consistent with
the calculated energy differences between dimerized phases with
different dimer long range ordering. Upon quick cooling to 50$^\circ$, the long-range
dimer ordering was restored 
suggesting that the high-temperature phase is a valence bond
liquid (not a glass), where dimerization occurs dynamically on a time scale
long compared to the characteristic time scale of our X-ray
measurements. While the concept of quantum spin liquid of dynamically
disordered spin-singlets is well known,
its classical analog, a
liquid of valence bonds dynamically disordered due to thermal
fluctuations, as it is our case, has been less investigated~\cite{Lakkis1976}.
Statistical physics of such an object should be nontrivial, bearing
resemblance to Maier-Saupe transitions in liquid crystals
and solid hydrogen\cite{H}. We hope that our results
will stimulate further experimental and theoretical studies in this
direction.

We thank M. Di Michiel and G. Khalliulin  for useful discussions and
the ESRF for support.  I.I.M.  acknowledges support from the Funding
from the Office of Naval Research (ONR) through the Naval Research
Laboratory's Basic Research Program. M.A., H.O.J. and R.V. thank the
German Science Foundation (DFG) for funding through SFB/TRR 49 and FOR
1346. S.S.V. thanks the Russian Foundation for Basic Research and the
Ministry of education and science of Russia for research programs
RFFI-13-02-00374 and MK-3443.2013.2. The work of D.Kh. was supported
by the DFG via FOR 1346 and Research grant 1484/2-1, and by Cologne
university via the German excellence initiative.


\begin{thebibliography}{99}
\bibitem{comment_bonds} \bibinfo{author}{ Note that the measured short bond is shorter than
the Ru-Ru metallic bond distance of 2.65 \AA.}
\end{thebibliography}
\end{document}